\documentclass[twocolumn,floatfix,prb,aps]{revtex4-1}
\usepackage{graphicx,url,color,nicefrac,color,multirow,amsmath,amssymb,amsfonts,graphicx,tabularx,amsfonts}
\usepackage{float}
\usepackage{placeins}
\usepackage[unicode=true,colorlinks=true]{hyperref}
\hypersetup{linkcolor=blue,citecolor=blue,urlcolor=blue}

\begin{document}

\title{Scaling of Loschmidt echo in boundary driven critical $\mathbb{Z}_3$ Potts model}
\author{Naveen Nishad, G J Sreejith}
\affiliation{Indian Institute of Science Education and Research, Pune 411008 India}
\date{\today}
\begin{abstract}
Low frequency perturbations at the boundary of critical quantum chains can be understood in terms of the sequence of boundary conditions imposed by them, as has been previously demonstrated in the Ising and related fermion models. Using extensive numerical simulations, we explore the scaling behavior of the Loschmidt echo under longitudinal field perturbations at the boundary of a critical $\mathbb{Z}_3$ Potts model. We show that at times much larger than the relaxation time after a boundary quench, the Loschmidt-echo has a power-law scaling with time as expected from interpreting the quench as insertion of boundary condition changing operators. Similar scaling is observed as a function of time-period under a low frequency square-wave pulse. We present numerical evidence which indicate that under a sinusoidal or triangular pulse, scaling with time period is modified by Kibble-Zurek mechanism, again similar to the case of the Ising model. Results confirm the validity, beyond the Ising model, of the treatment of the boundary perturbations in terms of the effect on boundary conditions.
\end{abstract}

\maketitle

\section{Introduction\label{sec:intro}}
The search for robust phenomena in interacting many body quantum systems far out of equilibrium has seen a recent surge of activity motivated by the increasing ability to probe quantum dynamics in artificial many body systems \cite{RevModPhys.86.153,bloch2012quantum}, and by the improvement in computational techniques that can reliably simulate dynamics in very large quantum many body systems\cite{PhysRevLett.93.076401,PhysRevLett.93.040502,Schollwock2011,PhysRevB.94.165116}. Attempts at developing general guiding principles like in equilibrium systems have prompted the investigation of tractable models and protocols for out of equilibrium systems.\cite{Moessner2017,Bukov}.
Systems out of equilibrium due to time-periodic Hamiltonians are among the simplest of such tractable systems. The constant stroboscopic unitary time evolution operators in such systems allow natural extensions of equilibrium notions such as steady state ensembles and quasienergies \cite{PhysRev.138.B979,PhysRevA.7.2203,PhysRevA.7.2203}, while at the same time demonstrating dynamical aspects that differ from equilibrium systems. Creative applications of Floquet physics\cite{Eckardt_2015,PhysRevB.78.235124} have already lead to the demonstration or prediction of many out of equilibrium phenomena including Floquet localization\cite{PhysRevLett.114.140401,PhysRevLett.114.140401}, freezing \cite{PhysRevB.82.172402,PhysRevB.90.174407}, time crystals \cite{choi2017observation,Zhang2017,PhysRevLett.117.090402,PhysRevLett.120.180602,PhysRevLett.120.180603}, topological phases \cite{PhysRevB.93.201103,PhysRevB.94.125105,potter2016classification,PhysRevB.93.245145,PhysRevB.96.155118,PhysRevLett.119.123601}, non adiabatic charge pumping\cite{PhysRevX.6.021013}, Floquet edge modes\cite{PhysRevX.3.031005,PhysRevB.88.155133,PhysRevB.94.045127} etc. Periodic driving in generic interacting systems however leads to heating\cite{PhysRevX.4.041048,PhysRevE.90.012110}, taking the system ultimately to a featureless time steady state. This fate can nevertheless be delayed in certain systems with long prethermalization times, strong disorder etc.\cite{PONTE2015196,PhysRevLett.116.250401,PhysRevX.7.011026,roeck2019slow}

A periodic drive wherein the time dependent part of the Hamiltonian is local could also lead to interesting physics in the long time limit, heating being avoided here due to dissipation into the surrounding medium. Lack of translational invariance as well as a need for large system sizes make analytical and numerical studies of such systems difficult. Local periodic drives at the boundary of critical 1D semi-infinite quantum system are nevertheless tractable in the long-time limit, owing to the possibility of mapping the problem to scenarios in 2D classical boundary critical phenomena\cite{CARDY2006333}. From the point of view of numerical experiments, long chains can be used to mimic semi-infinite systems allowing simulations of the system sufficiently long before finite size corrections appear, allowing access to long time scaling properties.

In this spirit, critical quantum transverse field Ising model subjected to a periodic longitudinal field (of amplitude $h_b$) was explored in Ref-\onlinecite{PhysRevLett.118.260602}. It was demonstrated that at low frequencies ($\omega=2\pi/T$, where $T$ much larger than the relevant relaxation times) the Loschmidt echo $|\left\langle \psi(0)|\psi(NT) \right\rangle|^2$ after $N$ time steps has a frequency dependence of the form $\left(\omega h^{-\nu}\right)^{N\gamma}$. When the time dependent boundary field has a square waveform, the exponent $\gamma=4h_{\text{BCC}}$ where $h_{\text{BCC}}$ is the scaling dimension of the boundary condition changing operator corresponding to the change of boundary condition (for example from a up-spin to down-spin). This exponent $\gamma$ is corrected by Kibble Zurek mechanism\cite{PhysRevLett.95.105701,PhysRevLett.101.016806,RevModPhys.83.863} to $\nicefrac{4h_{\rm BCC}}{1+\nu}$ when the boundary field has a triangular or sinusoidal waveform. Numerical simulations of the Ising model with integrability breaking perturbations suggested that the scaling is robust in the presence of interactions. Similar scaling was also observed in fermionic models related to the Ising model\cite{vasseur2014universal}.

In this work, we explore the quantum critical $\mathbb{Z}_3$ Potts model under similar boundary perturbations. The $\mathbb{Z}_3$ quantum Potts model generalizes the Ising model, and has a conformal critical point at the transverse field induced transition from the $\mathbb{Z}_3$ ordered phase to the paramagnetic phase. The Ising model following a Jordan Wigner transformation maps to a fermion problem\cite{PFEUTY197079} with a quadratic Hamiltonian and linear, exactly-solvable Heisenberg equations of motion allowing extensive analytical and numerical studies of its dynamics. The Potts model under a similar transformation, maps to a parafermion model with a quadratic Hamiltonian, which nevertheless does not yield linear equations of motion\cite{FRADKIN19801,PhysRevD.24.1562,HOWES1983169,Fendley_2012}. We therefore rely on matrix product states(MPS) time evolution \cite{PhysRevB.94.165116} to simulate the Potts chain and explore the scaling properties of Loschmidt echo.

The paper is structured as follows. In section \ref{sec:model} we introduce the Potts model, describe the protocol for periodic drive and summarize the scaling relations expected. Details of the numerical simulations are presented in section \ref{sec:numerics}. Results of the numerical simulation and their discussion are presented in section \ref{sec:results}.

\section{Model and description of scaling\label{sec:model}}

\begin{figure}
\includegraphics[width=\columnwidth]{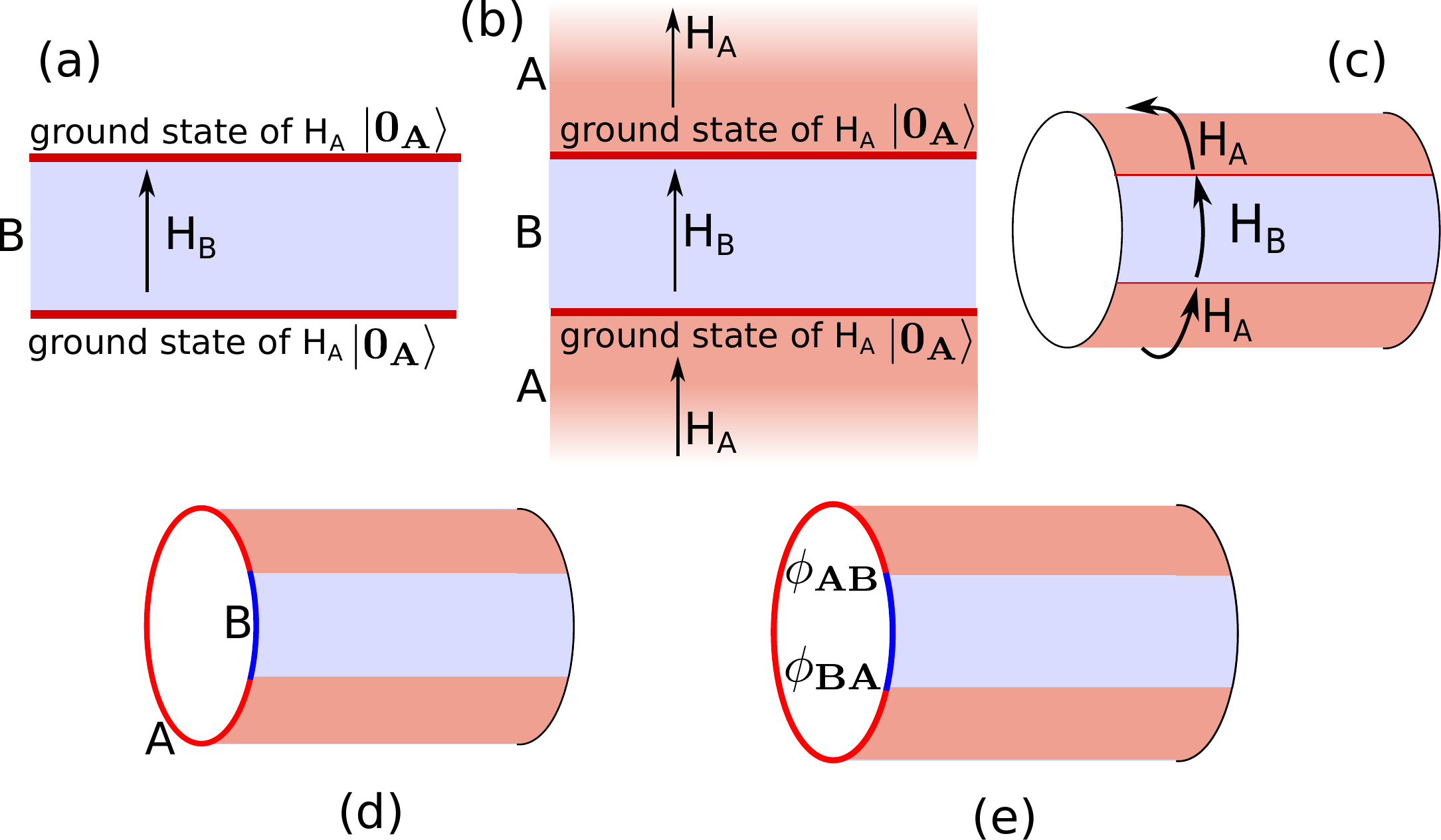}
\caption{(a) Overlap of the time evolved state $\left |\psi(t) \right\rangle=e^{-\imath H_Bt}\left|0_A\right\rangle$ with the initial state $\left|0_A \right\rangle$ shown schematically. In complex time, this is the partition function on a strip with the fixed state $\left|0_A\right\rangle$ at the top and bottom ends of the strip. (b) Since $\left|0_A\right\rangle$ is the ground state of $H_A$, this can be interpreted as the state evolved after a long time from an arbitrary state under complex time evolution under $H_A$. (c) Identifying the arbitrary initial and final states maps the overlap to a partition function on a cylinder with Hamiltonians $H_B$ and $H_A$ in two regions. (d) Effect of the two Hamiltonians can be approximated as enforcing two boundary conditions in the corresponding regions. (e) The changing boundary conditions can be interpreted as insertion of suitable boundary condition changing operators.
\label{fig:schematic}}
\end{figure}
In this section, we describe the Potts model and summarize the results that we borrow from boundary conformal field theory. The arguments, following Ref \onlinecite{PhysRevLett.118.260602} for the scaling behavior of the Loschmidt echo specialized to the scenarios considered in this paper are also presented.
The $\mathbb{Z}_3$ Potts model is described by the Hamiltonian
\begin{equation}
H_0 = -f\sum_i\kappa_i-J\sum_i\sigma_i\sigma_{i+1}^{\dagger}+ h.c \label{eq:ham}
\end{equation}
$\sigma$ and $\kappa$ are $\mathbb{Z}_3$ generalizations of the Pauli matrices $\sigma^z$ and $\sigma^x$ given by 
\begin{equation}
\sigma=
\begin{pmatrix}
1 & 0 & 0\\
0 & \omega & 0 \\
0 & 0 & \omega^2
\end{pmatrix},
\:\kappa=
\begin{pmatrix}
0 & 1 & 0\\
0 & 0 & 1\\
1 & 0 & 0
\end{pmatrix}
\end{equation}
where $\omega$ is $\exp(\frac{2\pi}{3}\imath)$. $\sigma$ and $\kappa$ satisfies the algebra $\sigma^3=\kappa^3=1$ and $\kappa \sigma=\omega\sigma \kappa $ at each site, generalizing the algebra of Pauli matrices. The Hamiltonian Eq \ref{eq:ham} is $\mathbb{Z}_3$ symmetric as it commutes with the generalized parity operator $P=\prod_{i=1}^N\kappa_i$. There is a continuous phase transition at $f=J$ from a three fold symmetry broken ordered phase ($f<J$) to a paramagnetic phase $f>J$. The transition point is a conformal critical point, with the operators in the model related to the $\mathcal{M}(6,5)$ minimal model and a non-diagonal modular invariant as the partition function.\cite{DiFrancesco:639405,Zamolodchikov:1985wn}

In this work, we study the Potts model with a boundary field. The Hamiltonian of the chain is time-dependent due to the boundary field and has the following form:
\begin{equation}
H = -h(t)m + H_0 \text{ where }m=\frac{\sigma_0+\sigma_0^\dagger}{2}\label{eq:drivenham}
\end{equation}
For a positive $h(t)$, the ground state of the system has the boundary spin pointing in the $\left \langle \sigma_0\right \rangle \sim 1$ state (fixed boundary condition) and for negative $h(t)$, the boundary spin in the ground state is in a mixed state of $\langle\sigma_0\rangle=\omega$ and $\langle\sigma_0\rangle=\omega^2$ (mixed boundary condition). Two other fixed and mixed boundary conditions are obtained if the boundary term is replaced with $h\sigma_0+h.c.$ and ${\arg}(h)$ is $\pm 2\pi/3$ and $\pi\pm 2\pi/3$ respectively. The fixed boundary conditions are RG fixed points in the parameter plane of the complex boundary field. Mixed boundary conditions under perturbation of $h$ flow into nearby fixed boundary condition points.\cite{Affleck_1998} 

We consider a scenario where a time-periodic boundary field $h(t)$ toggles between positive and negative values. Initially, the chain is assumed to be in the ground state of the initial Hamiltonian. We will study the fate of the Loschmidt echo defined as the overlap of the time evolved state with the initial state $\mathcal{L}=|\left \langle \psi(0)|\psi (t) \right \rangle|^2$. 

Consider the scenario where the boundary field is quenched from $+h_b$ to $-h_b$ ($h_b>0$). The Loschmidt echo after this quench is given by $\mathcal{L}(t)=
|\left \langle 0_A |\exp({-\imath t H_B})|0_A\right\rangle|^2$ where $\left | 0_A \right \rangle$ is the ground state of the initial Hamiltonian $H_A=H_0-h_bm$, and $H_B$ is $H_0+h_bm$. In order to map the problem to a 2D classical system, we can analytically continue to the complex time. The initial state upto proportionality constants can be identified with $\lim_{s \to \infty} \exp\left(-H_A s \right)\left | \alpha \right \rangle$ where $\alpha$ is a generic state with non-zero overlap with the ground state (schematically shown in Fig \ref{fig:schematic}). Summing over all such states $\alpha$ maps the Loschmidt echo to 
\begin{equation}
\mathcal{L}(-\imath \tau) \sim 
\lim_{s\to \infty}
{\rm Tr}
\left[
e^{-sH_A}
e^{-\tau H_B}
e^{-sH_A}
\right]
\end{equation}
This can be identified with the partition function of a 2D classical system with periodic boundary conditions (Fig \ref{fig:schematic} c) and with a boundary field $-h_b$ along a tiny duration/distance along the imaginary time of $\tau$, and $+h_b$ otherwise. Following the idea introduced in Ref-\onlinecite{PhysRevLett.118.260602} we can approximate the effect of the boundary field to be to pin the boundary condition. Accordingly, the partition function can be replaced with that of a system with fixed boundary condition (labelled A in Fig \ref{fig:schematic}d) whenever $h>0$ and mixed boundary condition whenever $h<0$ (labelled B in Fig \ref{fig:schematic}d). A key result from boundary conformal field theory is that such a change in boundary condition can be interpreted as insertion  (Fig \ref{fig:schematic}e) of certain boundary condition changing operators.\cite{cardy1989boundary,CARDY2006333,DiFrancesco:639405} The partition function in Fig \ref{fig:schematic}d is interpreted as the correlation function of these operators separated by a distance $\tau$. The dominant scaling dimension $h_{\rm BCC}$ of the operator that changes the boundary condition between the free and mixed boundary conditions can be inferred to be $2/5$. The Loschmidt echo $\mathcal{L}(\imath \tau)$ scales in the same manner as the square of the correlation function at a distance along the boundary $\tau$ giving the result $\mathcal{L}(t)\sim|t|^{-4h_{\rm BCC}}$.

The above approximation is valid once the boundary spins have relaxed (over a time scale $h_b^{-\nu z}$ where $z=1$ for the Potts model and $\nu$ is the boundary field correlation length exponent) in response to the boundary field quench. Treating this as the short time scale in the problem, the Loschmidt echo scales as $\mathcal{L}(t)\sim |t h_b^\nu|^{-4h_{\rm BCC}}$. For the Potts model, it was argued in Ref-\onlinecite{Cardy1984} that the spin-spin correlation function scales with distance as $1/r^{\eta_{||}}$ where $\eta_{||}=\frac{4}{3}$. Comparing this with $2d-2/\nu$ (for $d=1$ dimensional boundary), we infer the correlation length exponent along the boundary to be $\nu=3$.  Loschmidt echo under square-wave boundary field (amplitude $h_b$, period $T$) will now be an $N$ point correlation function which scales with $T$ as $\sim(Th^\nu)^{-4Nh_{\rm BCC}}$. Note that the prefactors of the relation will have an $N$ dependence in general which is not captured by these arguments, though this was absent in the Ising model. 

For a periodic perturbation such as a sinusoidal or triangular wave that crosses $h=0$ at a finite slope, unlike the square-wave, the short time scale is replaced by the Kibble Zurek time scale $\sim \lambda^{-\frac{z\nu}{1+z\nu}}$ ($z=1$ for our system) where $\lambda \sim h_b/T$ is the rate at which the boundary field crosses the gapless system appearing at $h(t)=0$. With this the Loschmidt echo scales with time as $\sim (Th_b^\nu)^{-4N \frac{h_{\rm BCC}}{1+\nu}}$. 

In summary, for the cases studied here, after a quench from $h=+h_b$ to $-h_b$ the Loschmidt echo should scale with time as $\mathcal{L}(t)\sim(th_b^3)^{-\gamma}$ where $\gamma=8/5$. Loschmidt echo scales with the time period as $\mathcal{L}(t)\sim(Th_b^3)^{-\gamma N}$ where $\gamma=8/5$ for square wave and $\gamma=2/5$ for a triangular or sinusoidal wave.

\begin{figure}
\includegraphics[width=\columnwidth]{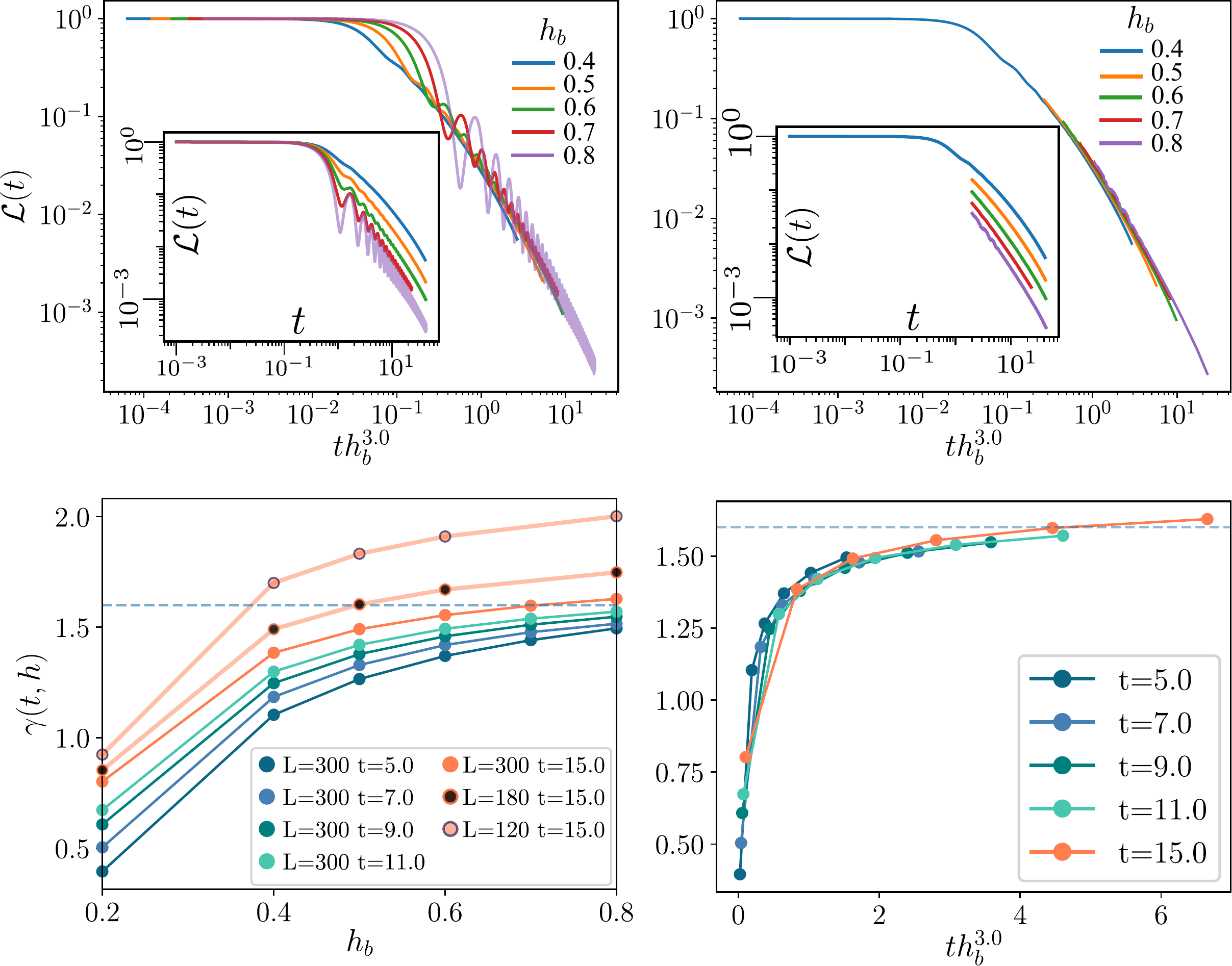}
\caption{(a) Loschmidt echo after a quench of the boundary field from $h_b$ to $-h_b$ plotted as a function of the rescaled time $t/\tau=th_b^3$ for system size L=300. Inset shows the same data as a function of time. (b) Same as panel (a) but after filtering out the oscillatory components of the Loschmidt echo. (c) Exponent $\gamma(t,h_b)$ estimated near specific times $t$ plotted as a function of $h_b$. $\gamma(t,h_b)$ is estimated from the slope of $\log \mathcal{L}(t)$ vs $\log t$ shown in panel (b). Comparison with data from $L=120,180$ indicate saturation to the expected exponent $8/5$ at large $Th^{3}$ in large systems.
(d) Same as panel (c) but plotted as a function of rescaled time.
\label{fig:quench}}
\end{figure}

\section{Numerical methods\label{sec:numerics}}
We use a matrix product representation of the states and Trotter decomposed time dependent unitary time evolution operators represented as matrix product operators to calculate the Loschmidt echo.\cite{Schollwock2011} Ground state calculation was done using DMRG implementation in MPS language. The time evolution was performed in steps of $\Delta t=10^{-3}$. The unitary time evolution was implemented as a sequence of two site gates using a time-dependent Suzuki-Trotter fourth order approximant \cite{Hatano2005}. The bond dimension of $300$ was used for the calculations presented. Calculations were performed with bond-dimensions $150$ and $100$ in addition, to ensure that the Loschmidt echo curves (upto $t=26/J$) have converged (comparison of bond dimensions is presented in the Appendix \ref{app:bond}). In general the required bond dimensions were higher than $300$ for simulations with $h_b<0.2$ i.e. closer to the critical system, and therefore we have relied on data from $h_b\geq0.4$. $J=h=1$ was used for all the calculations presented. Chains of size $L=180,220,260$ and $300$ were studied and we have presented the data for the largest system size $L=300$. Finite size effects sets in as the wavefronts from the change in boundary condition propogates into the chain. In order to avoid finite size effects, time evolution only for $N=1$ could be reliably verified. In this work, we have restricted to the investigation of boundary fields that cause changes between fixed and mixed boundary conditions. Due to large bond dimensions, study of quenches into or from free boundary condition ($h=0$) is unreliable. While the expected scaling dimenion $h_{\rm BCC}=1/4$ was accurately obtained in those simulations, correlation length exponent $\nu$ which determines relaxation time scale showed significant system-size and bond-dimension dependence.
\section{Numerical results\label{sec:results}}

\begin{figure}
\includegraphics[width=\columnwidth]{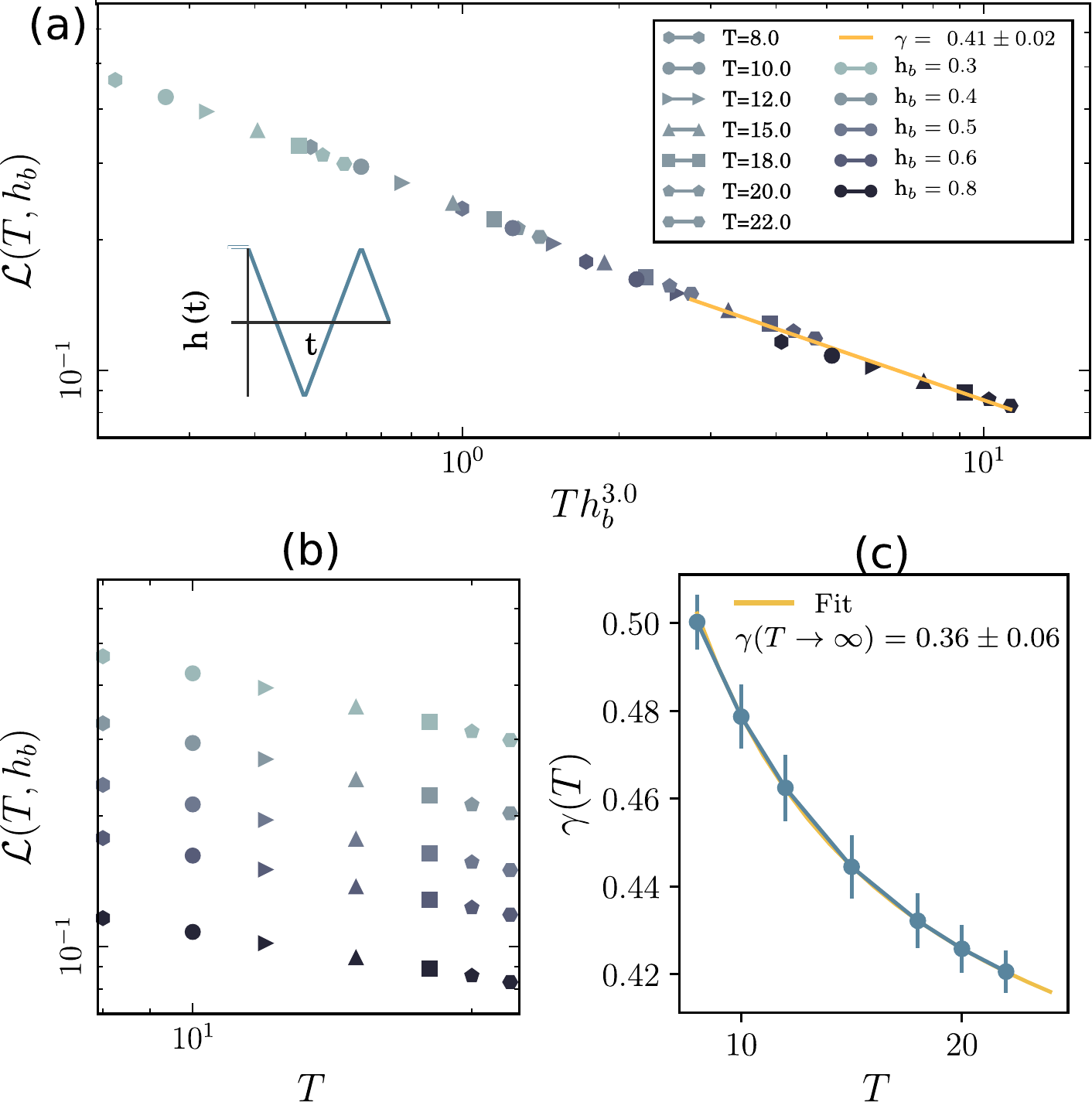}
\caption{ (a) Loschmidt echo $\mathcal{L}(T)$  at the end of a time period plotted as a function of rescaled time period $T/\tau=Th_b^{3}$ for a triangular-wave boundary field for system size L=300. (b) $\mathcal{L}(T)$ plotted as a function of the time period. (c) Scaling exponent $\gamma (T)$ estimated using data at fixed $T$ by calculating the slope of $\ln \mathcal{L}$ vs $\ln Th_b^3$. Large $T$ limit of $\gamma$ is obtained by fitting the data to $\gamma(T)=\gamma(\infty)+\frac{axT^{-x}}{1+aT^{-x}}$\label{fig:triangular}}
\end{figure}

In this section we present the numerical results for dynamics of Loschmidt echo under boundary perturbations. Figure \ref{fig:quench} summarizes the results for Loschmidt echo as a function of time $t$ after a quench from $H_+=H_0-h_bm$ to $H_-=H_0+h_b m$, with the system starting from the ground state of the initial Hamiltonian. Results are presented for the largest system studied ($L=300$). The Loschmidt echo (inset-a) approaches a power law scaling for $t$ larger than the relaxation time scale $\tau=h_b^{-3}$, however for very large $t$, finite size effect sets in leading to deviation from power-law scaling behavior. For the small $h_b\leq0.4$, the relaxation time is large and comparable to the time scale for onset of finite size effects, as a result a scaling region is not evident in this case. A power-law regime emerges for larger boundary fields where $\tau$ is smaller i.e. when $h_b$ increases. However, as the boundary field approaches the local energy scale of $J=1$, the power-law gets masked by an oscillatory component. Empirically we find that these oscillations decay with time and have a constant frequency. Figure \ref{fig:quench}(b-inset) shows the same data after filtering out a oscillatory component, and reveals a power-law scaling in the backdrop of the oscillations (Details of the filtering are described in Appendix \ref{app:fft}). The Loschmidt echo trace $\mathcal{L}(t)$ for different $h_b$ shows a scaling collapse when the time $t$ is rescaled by the relaxation time scale (Fig \ref{fig:quench}(a)). Similar collapse is also seen in the Loschmidt echo after filtering out the oscillatory components.

Figure \ref{fig:quench}(c) shows the negative of the slope in the log-log plot of the filtered data as a function of $h_b$ at different fixed times $t$. The slopes approach $\gamma=8/5$ with increasing $t$ as well as with increasing $h_b$ (decreasing $\tau$). Empirically, we find that the slopes approach the saturation values as a function of $th_b^3$ (Fig \ref{fig:quench} (d)). For small systems, finite size effect sets in before the slope saturates. This can be seen from the slopes for $L=120,180,300$ at $t=15$ shown in Fig \ref{fig:quench} (c). With increasing $L$, the time-scale when finite size effect sets in increases and saturation of the slope is evident.  

Under a square-wave boundary perturbation oscillating between $H_{\pm}=H_0+\mp h_bm$, when the initial state is the ground state of the initial Hamiltonian $H_+$,  the 
Loschmidt echo after a single time period $T$ namely $|\left \langle \psi(0)| e^{-\imath H_+ T/2}e^{-\imath H_- T/2} | \psi(0)\right \rangle|^2$ is the same as the Loschmidt echo after a time $T/2$ following a quench. Thus $\mathcal{L}(T,h_b)\sim (Th_b)^{-\gamma}$ where $\gamma=8/5$.

Figure \ref{fig:triangular} summarizes the results for the Loschmidt-echo under a triangular wave like boundary perturbation after a single time-period. The initial state is the ground state of the Hamiltonian $H=H_0-h_b m$. The Hamiltonian is linearly changed to $H=H_0+h_b m$ and then linearly back to $H=H_0-h_b m$ over a total time $T$. Figure \ref{fig:triangular}(b) shows the Loschmidt echo for different values of $T$ and $h_b$. The data at different values of $T$ and $h_b$ collapse when expressed as a function of $Th_b^3$ (Figure \ref{fig:triangular}(a)). The slope estimated from the largest values of $Th_b^3$ was $0.41\pm 0.02$. Exponent $\gamma$ obtained by fitting the data to $\mathcal{L}(T,h_b)=A(Th_b^3)^{-\gamma}$ separately for each value of $T$ is shown in Fig \ref{fig:triangular}(c). $\gamma(T)$ approaches $0.4$ as $T$ increases.  To get the estimate of the asymptotic value of the slope, we fitted the data to $\gamma(T)\approx\gamma(\infty)+cT^{-d}$ (Appendix \ref{app:estgamma}) to get $\gamma(\infty)=0.36\pm0.06$. These values are consistant with the expected value of $\gamma=\frac{4h_{\rm BCC}}{1+\nu}=\frac{2}{5}$.

\begin{figure}
\includegraphics[width=\columnwidth]{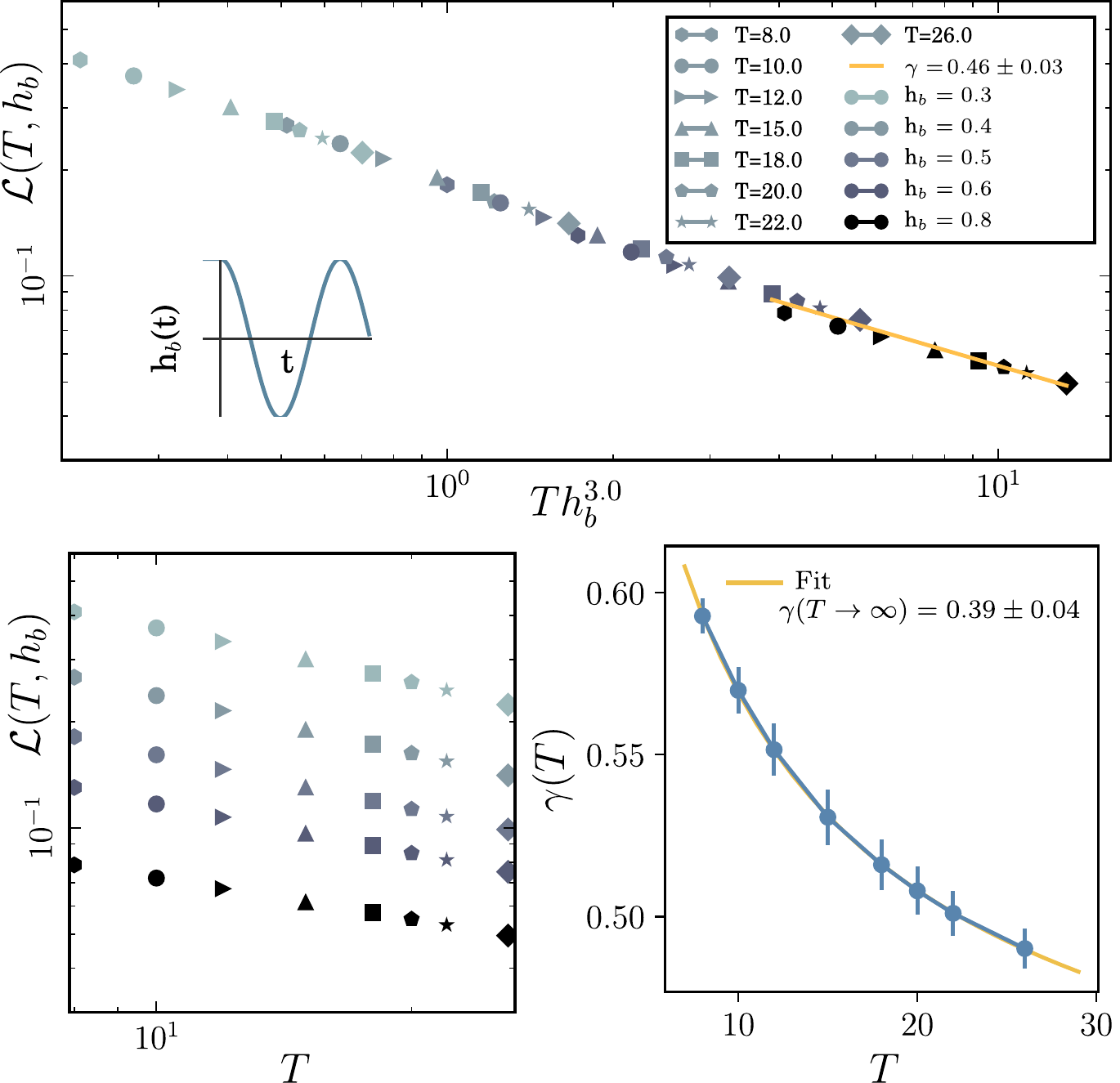}
\caption{ Loschmidt echo for a sinusoidal wave perturbation for system size L=300. Loschmidt echo as a function of rescaled time period $T/\tau=Th_b^3$ is shown in panel (a). Panel (b) shows the same data as a function of $T$. Scaling exponent $\gamma(T)$ estimated from data at fixed time period $T$. Asymptotic value is estimated by fitting to $\gamma(\infty)+\frac{axT^{-x}}{1+aT^{-x}}$. \label{fig:sin}}
\end{figure}

Figure \ref{fig:sin} shows the results for the Loschmidt-echo under a sinusoidal wave boundary perturbation after a single time-period $T$. The initial state is again the ground state of the Hamiltonian $H=H_0-h_b m$. The Hamiltonian is changed to $H=H_0-h_b m$ and then back to $H=H_0+h_b m$ in a sinusoidal manner. The data at different values of $T$ and $h_b$ (Fig \ref{fig:sin}(b)) collapse when expressed as a function of $Th_b^3$ (Figure \ref{fig:sin}(a)). Slope estimated from the largest values of $Th_b^3$ was $0.46\pm 0.03$. We believe the deviation is due to $T$ not being large enough. Exponent $\gamma$ obtained by fitting the data at each fixed $T$ to $\mathcal{L}(T,h_b)=A(Th_b^3)^{-\gamma}$ again drifts towards $0.4$ at large $T$. From fitting this data to $\gamma(T)\equiv -\frac{d\ln \mathcal{L}}{d\ln T}\approx\gamma(\infty)+ cT^{-d}$, $\gamma(\infty)$ was found to be $0.39\pm 0.04$, which is again consistant with the expected value of $2/5$ (details of fitting in the appendix \ref{app:estgamma}).

\section{Conclusions\label{sec:conclusion}}
We have presented the results of dynamics of Loschmidt echo from numerical experiments on low frequency boundary fields on large, but finite, quantum critical $\mathbb{Z}_3$ Potts chains. The results obtained after single time period are in close agreement with what is expected from interpreting the fields as simply imposing a corresponding set of boundary conditions: Loschmidt echo scales with frequency as $(\omega h_b^{-\nu})^{\gamma}$ where $\gamma$ is determined by (i) the boundary condition changing operators corresponding to the sequence of boundary condition changes imposed by the boundary field and (ii) the manner in which the boundary field crosses the $h=0$ point. Unlike most results on Floquet systems that hold in the high-frequency regime derived from some variant of the Magnus expansion, the scaling results explored here hold in the lowest frequency regime. It will be very useful to extend the formalism beyond Loschmidt echo and to physically measurable correlations of energy and magnetization.

Scaling of $\mathcal{L}$ in the model studied here was tractable due to the mapping to the classical 2D conformal critical system. The Potts model can be generalized to a broader class of chiral $\mathbb{Z}_3$ symmetric models\cite{PhysRevB.24.398,PhysRevB.24.5180,Fendley_2012} with $z\neq 1$, lacking a simple mapping to a 2D classical conformal critical point.\cite{PhysRevA.98.023614,PhysRevB.98.205118,PhysRevB.99.184104,PhysRevB.97.014309} It will be interesting to explore whether similar scaling properties apply to boundary field quenches within these models and if they do what dictates the scaling exponents? 

\acknowledgments
We thank Manoranjan Kumar and Hitesh Changlani for useful discussions. Calculations were performed using the codes built on ITensor Library\cite{ITensor}. Part of the calculations were performed using computing resources acquired under the DST/SERB grant ECR/2018/001781.
\bibliography{biblio_Potts}
\bibliographystyle{apsrev}

\appendix
\section{Removing oscillatory part of $\mathcal{L}(t)$\label{app:fft}}
\begin{figure}[H]
\includegraphics[width=\columnwidth]{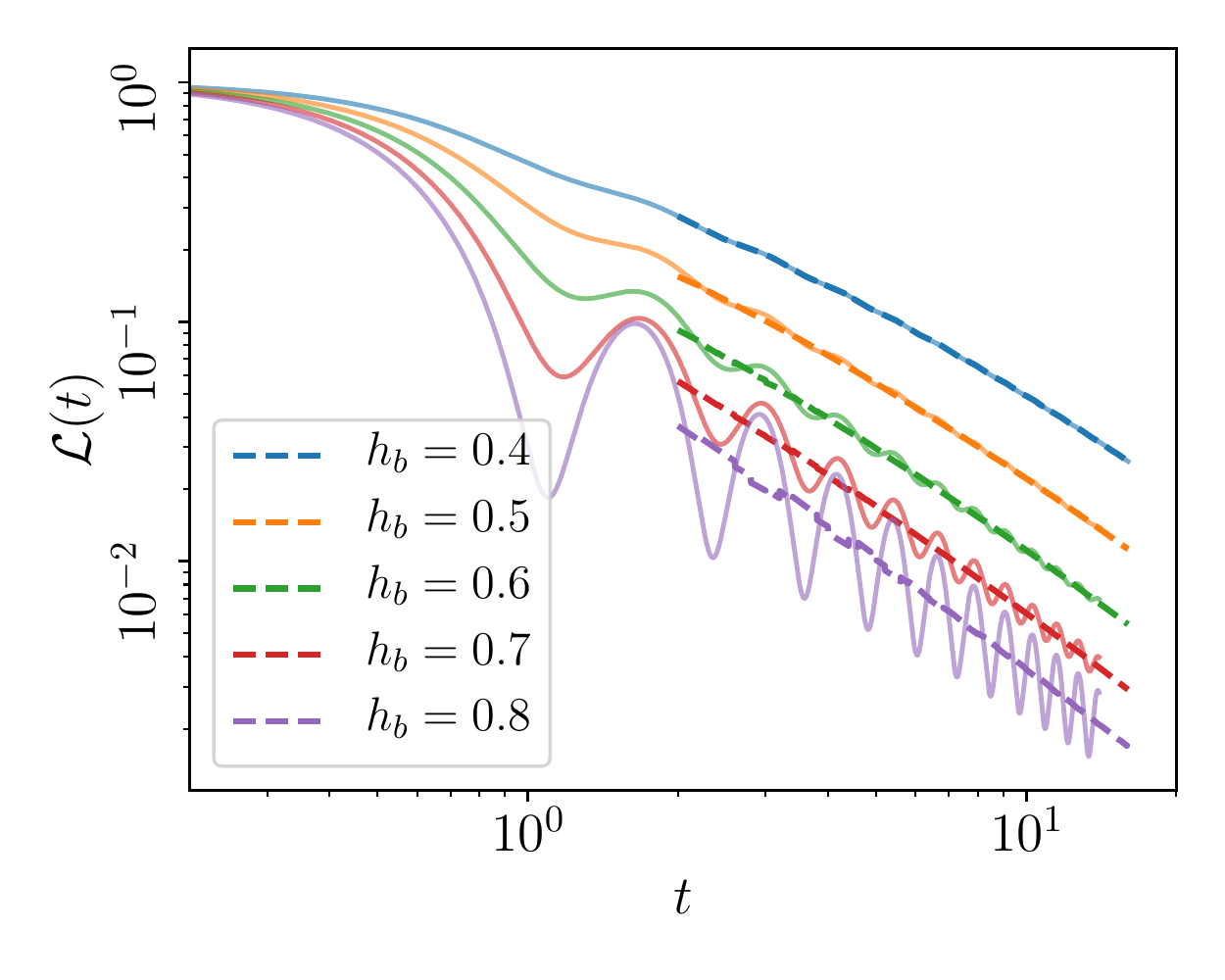}
\caption{Continuous lines show the actual Loschmidt echo as a function of time. Dashed lines show the Loschmidt echo after filtering out the oscillations\label{fig:filtered}} 
\end{figure}

Loschmidt echo after a quench of the boundary field from $h_b$ to $-h_b$ shows oscillatory behavior for $h_b\sim J$. The amplitude of these oscillations decay with time and thererfore should be absent at large times. In order to access the underlying scaling form within the finite time data accessible using numerical simulations, we filter out the oscillatory part of the data. Filtering of the data is primarily useful for $h_b=0.8$ (Fig \ref{fig:filtered}).
Here we describe the scheme used for filtering out the oscillations. For each value of $t$, the data $\ln\mathcal{L}$ in a narrow window (of the order of a two wavelength in the plot) near $t$ was fitted to the following function.
\begin{equation}
A\text{sin}(\omega t+\phi)+B+r \ln(t)\nonumber
\end{equation}
The filtered data at time $t$ was then taken to be $B+r \ln(t)$. Figure \ref{fig:filtered} shows the Loschmidt echo before and after filtering out the oscillations.

\FloatBarrier
\section{Bond dimension \label{app:bond}}
\begin{figure}[H]
\includegraphics[width=\columnwidth]{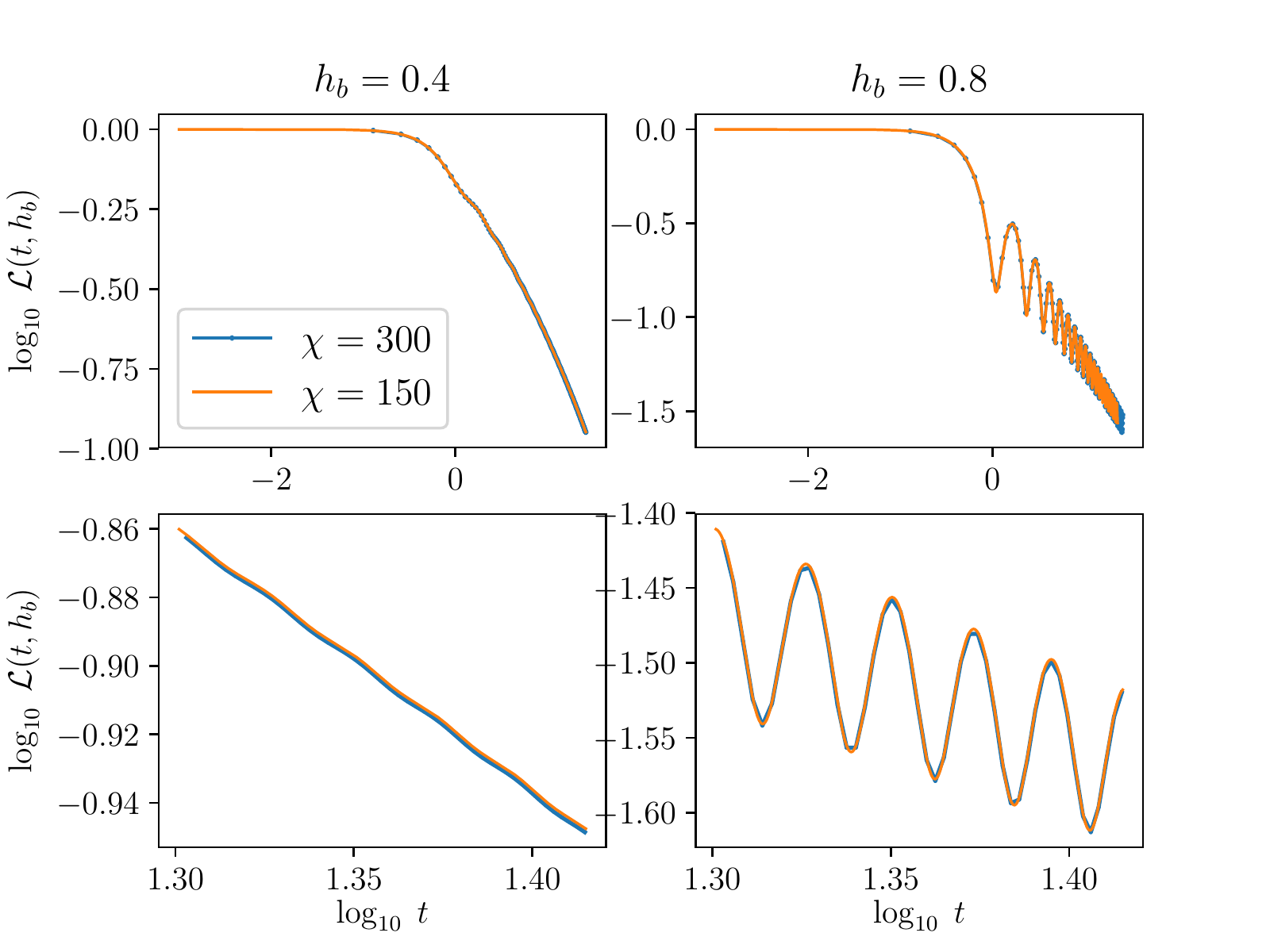}
\caption{Loschmidt echo from calculations with bond dimension $\chi=150$ (orange) and $\chi=300$(blue with markers) for two different values of $h_b$ in a system of size $L=300$. panels c and d show the zoomed in plots in near the large $t$ end of the data in panels a and b. \label{fig:bonddim}} 
\end{figure}

Here we consider the bond dimension ($\chi$) dependence of the Loschmidt echo. Fig \ref{fig:bonddim}(a,b) show the Loschmidt echo as a function of the rescaled time $Th_b^3$ for $h_b=0.4$ and $h_b=0.8$ respectively. Data obtained from calculations with bond dimension $\chi=150$ (orange) and $\chi=300$ (blue with markers) are both shown. There is no appreciable difference in the results from the two calculations for the time scales studied here.

\section{Estimation of $\gamma(\infty)$\label{app:estgamma}}
In this section we describe the fitting used to estimate $\gamma(T\to\infty)$ from data in Fig \ref{fig:triangular}(c) and \ref{fig:sin}(c).
We assume that Loschmidt echo has a subleading correction to scaling with $T$, i.e. a form $\mathcal{L}\sim T^{-\gamma(\infty)}(1+aT^{-x})$. 
From this form the, power-law estimated from data at fixed $T$ can be estimated to be 
$\gamma(T)\equiv -\frac{d \ln \mathcal{L}}{d\ln T} = \gamma(\infty)+\frac{axT^{-x}}{1+aT^{-x}}$. 
The $\gamma(T)$ vs $T$ data was fitted by maximizing log-likelihood assuming a simple normal-distribution of error. 
\end{document}